\documentclass[11pt]{article}

\usepackage{amsmath, amssymb, mathrsfs, bm, latexsym,graphics,color, graphicx,url,floatflt}
\usepackage[authoryear]{natbib}

\newcommand{\ignore}[1]{}

\newcommand{\blackslug}{\penalty 1000\hbox{
    \vrule height 8pt width .4pt\hskip -.4pt
    \vbox{\hrule width 8pt height .4pt\vskip -.4pt
          \vskip 8pt
      \vskip -.4pt\hrule width 8pt height .4pt}
    \hskip -3.9pt
    \vrule height 8pt width .4pt}}

\newcommand{\qed}{\hspace*{\fill}\blackslug}

\def\boxit#1{\vbox{\hrule\hbox{\vrule\kern4pt
  \vbox{\kern1pt#1\kern1pt}
\kern2pt\vrule}\hrule}}

\begin{document}

\title{From data towards knowledge: Revealing the architecture  of signaling systems
by unifying knowledge mining and data mining of systematic
perturbation data}

\author{Songjian Lu$^{\text{}1}$, Bo Jin$^{\text{}1}$, Ashley Cowart$^{\text{}2}$,
Xinghua Lu$^{\text{}1,*}$}


\maketitle

1. Department of Biomedical Informatics, University of Pittsburgh,
Pittsburgh, PA 15232.

2. Dept Biochemistry and Molecular Biology, Medical University of
South Carolina, Charleston, SC 29425.

$*$ Corresponding Author: Xinghua Lu, Department of Biomedical
Informatics, University of Pittsburgh,  5607 Baum Boulevard,
Pittsburgh, PA 15232.

\newpage

\section*{Abstract}

\paragraph*{}
Genetic and pharmacological perturbation experiments, such as
deleting a gene and monitoring gene expression responses,  are
powerful tools for studying cellular signal transduction pathways.
However, it remains a challenge to automatically derive knowledge
of a cellular signaling system at a conceptual level from
systematic perturbation-response data.  In this study, we explored
a framework that unifies knowledge mining and data mining
approaches towards the goal.  The framework consists of the
following automated processes: 1) applying an ontology-driven
knowledge mining approach to identify functional modules among the
genes responding to a perturbation in order to reveal potential
signals affected by the perturbation; 2) applying a graph-based
data mining approach to search for perturbations  that affect a
common signal with respect to a functional module, and 3)
revealing the architecture of a signaling system organize
signaling units into a hierarchy based on their relationships.
Applying this framework to a compendium of yeast
perturbation-response data, we have successfully recovered many
well-known signal  transduction pathways; in addition, our
analysis have led to many hypotheses regarding the yeast signal
transduction system; finally, our analysis automatically organized
perturbed genes as a graph reflecting the architect of the yeast
signaling system.  Importantly, this framework transformed
molecular findings from  a gene level to a  conceptual level,
which readily can be translated into computable knowledge in the
form of rules regarding the yeast signaling system, such as ``if
genes involved in MAPK signaling are perturbed,  genes involved in
pheromone responses will be differentially expressed''.
\newpage{}

\section*{Introduction}
Model organisms, such as \textit{Saccharomyces cerevisiae} and
\textit{Drosophila melanogaster}, are powerful systems to study
cellular signal transduction, because they are amenable to
systematic genetic and pharmacological  perturbations, enabling
biologists to infer whether a gene is involved in a signal
transduction pathway through studying perturbation-response data.
The premise for elucidating signal transduction pathways from
systematic perturbation  experiments is that, if perturbation of a
set of genes  consistently causes  a common cellular response,
e.g., a phenotype presented as the differential expression of a
module of genes, the perturbed genes  are likely the members (or
modulators) of the signal transduction pathway that leads to the
phenotype.

In this study, we refer to  a \textit{signal}  from an information
theory \citep{Cover2006} point of view, in which a signal is a
latent variable whose state contains information  with respect to
another variable, e.g., the expression state of a gene module or
the state of another signal.  From the same viewpoint, a signaling
system  consists of a set of  latent variables connected as a
network, in which an edge exists between a pair of signals if the
state of one signal affects that of the other, i.e., information
can be transmitted between the signals, and the relay of signals
along the paths in the network enables the system to encode
complex information.  From a cell biology viewpoint, a signal
transduction pathway consists of a collection of signaling
molecules that detect and transmit a signal that has a  physical
or chemical form, e.g., the represent of pheromone in the
environment.  In such a system,  a signal is encoded as a change
in the state of a signaling molecule, often manifested as a change
in the structural conformation of a protein, chemical modification
of a signaling molecule, or a change in the concentration of a
signaling molecule.  While it would be ideal to find a one-to-one
mapping between the signaling molecules in cells and the signals
in the information theory framework, such a mapping can be
difficult to obtain and too complex to represent.  Representing
cellular signaling systems within the abstract information-theory
framework provides the following advantages: 1) it enables us to
use  latent variables to represent the state of  yet unknown
signaling molecules; 2) it allows us to represent the biological
signals encoded by a group of signaling molecules into a
single-bit signal, if the signals encoded by these molecules
convey a common piece of information with respect to other
variables.  We refer to such a group of signaling molecules as  a
\textit{signaling unit}. The following example illustrate the
parallelism between the biological entities and and their
counterparts in a computational model.  A pheromone receptor in a
yeast cell and its associated G-proteins can be thought  of as one
signaling unit, as they function together to detect the  signal of
pheromone in an inseparable manner.  Another exemplary  signaling
unit is the cascade of mitogen-activated protein kinases  (MAPKs),
which transduce signals among themselves through a chain of
protein phosphorylation reactions almost in a deterministic
fashion.  The states of these signaling units can be represented
as two single-bit signals in a computational model.    When a
yeast cell is exposed to pheromone, the receptor unit detects the
signal and transmit the signal to MAPKs
unit\citep{Gustin,Herskowitz1995}, which further relays the signal
to down stream signaling units to regulate expression of
downstream genes involved in mating.  These relationships between
signaling units can be represented as edges in the model.
Moreover, in addition to  pheromone response, the MAPK signaling
unit also interacts with other signaling units to transmit the
signals that affect  filamentation/invasion processes
\citep{Gustin,Herskowitz1995}; such branching and cross-talks
between different signaling pathways can be represented as a
network connecting signals in the computational model.   Thus, the
general task of using systematic perturbation data to study a
cellular signaling system can be reduced to the following specific
tasks:  1) revealing the signals embedded in the convoluted
molecular phenotype data such as microarrays,  2) identifying
perturbed genes that affect a common signal,  3) grouping
perturbed genes into signaling units based on the information they
encode,  and 4) inferring the paths between signaling units where
a path may or may not correspond to a signal transduction pathway
in conventional cell biology.

In the seminal work by \cite{Hughes}, yeast cells were subjected
to over 300 types of systematic perturbations (gene deletions and
chemical treatments\footnote{From here on,  we refer to such a
treatment  experiment as a perturbation instance}) and the
transcriptional responses to the perturbations were measured using
microarrays.  This dataset has been widely used to test different
computational approaches for investigating the relationship
between perturbed genes  and responding genes
\citep{Hughes,TanaySAMBA2002,OurfaliSPINE2007,MarkowetzNEM2007,Yeger-Lotem,Huang}.
For example, using a conventional hierarchical clustering
approach, Hughes \textit{et al} grouped  perturbed genes into
clusters to elucidate the cellular functions of some genes, based
on the fact that perturbing these genes produced gene expression
profiles similar to those resulting from perturbing the known
members of certain pathways. To relax the requirement of global
similarity by  hierarchical clustering, other researchers have
studied approaches to connect a subset of perturbation instances
to a subset of  responding genes in order to find context specific
information between the perturbation and the
responses~\citep{TanaySAMBA2002}.  Such a  task is often  cast as
a   biclustering problem \citep{MadeiraBiClustering2004,
Cheng2000, Erten2010}.  More recently,  sophisticated  graph-based
algorithms have been applied to the dataset to study potential
signal pathways   \citep{Yeger-Lotem,Huang,OurfaliSPINE2007}. The
basic idea underlying the studies by Yeger-Lotem \textit{et al}
and Huang \textit{et al } is to model the information flow from
perturbed genes to responding genes through a PPI network by
employing graph search algorithms, e.g., price collecting Steiner
tree algorithms.

While the above studies have led to many biological insights
regarding the system at a gene level,  they did not address the
task of discovering signaling units and representing the findings
at a conceptual level in order to derive computable knowledge,
such as the rule:  \textit{if a gene involved in MAPK pathway  is
deleted, the cellular response to pheromone will be affected}.
Transforming experimental data into concepts  and further
elucidating the relationship among the concepts are critical steps
of knowledge acquisition and knowledge representation.  The scale
of contemporary biotechnologyies further necessitates
computational approaches to perform such tasks in an automated
manner in order to facilitate knowledge discovery by human
experts.   Yet,  the development of such techniques  is severely
lagging behind the pace of data generation.  In this paper, we
report  a proof of concept framework that unifies knowledge mining
and data mining to derive  knowledge regarding a signaling system
in an automatic manner, and we refer to the overall approach as
ontology-driven knowledge discovery of signaling pathways (OKDSP).
We  tested the framework using the yeast perturbation-response
data by \cite{Hughes} to illustrate its utility.

\section*{Results and Discussion}
A key step of ``reverse engineering''  signaling pathways using
systematic perturbations data is to identify perturbations that
convey the same information, in other words, to first find the
``jigsaw puzzle'' pieces belonging to a signal transduction
pathway.   For example, a classic yeast genetic approach  is to
search for deletion strains that exhibit  a  common phenotype as a
means for identifying genes potentially involved in a  signaling
pathway carrying information with respect to the phenotype
\citep{Winzeler1999}.   The advent of genome technologies enables
biologists to use genome-scale data, such as gene expression data,
as ``molecular phenotypes'' to study the impact of systematic
perturbations \citep{Hughes}.   In general,   a perturbation
treatment,  such as deleting a gene, often affects multiple
biological processes.  For example, deleting a gene involved in
ergosterol metabolism will affect the organization of cell
membrane, which in turn will affect multiple signaling pathways
located in the membrane.  As such, the overall cellular response
to a perturbation instance, which often manifested as a long list
of differentially expressed genes,  inevitably reflects a mixture
of responses to multiple  signals.  Thus, we are confronted with
two fundamental tasks when studying systematic perturbation data:
1) dissecting signals from the convoluted gene expression
responses to a perturbation instance, i.e., finding a module of
genes whose expression state reflects the state of a signal
transduced along a signaling pathway,  2) identifying a set of
perturbation instances that affects the signal regulating a common
expression module.

To address the tasks, we hypothesize that, if a module of
genes---whose functions are coherently related---responds to
multiple perturbation instances in a coordinated manner, the genes
in the module are likely regulated by a common signal, and the
perturbation instances  affect this signal.     Based on this
assumption, we can  first decompose the overall  expression
response to a perturbation instance into functional modules,  with
each module potentially responding to a distinct signal; then  we
can  investigate if a functional module is repeatedly affected in
multiple perturbation instances.   In this study, we developed an
ontology-based knowledge mining approach to identify functional
modules, and we then  developed a novel bipartite-graph-based data
mining approach to search for perturbation instances affecting a
common signal.  Based on the results from the steps above, we
further identified signaling units and revealed their organization
in a signaling system using a graph-based algorithm.

\subsection*{Identifying  functional modules through  knowledge mining}
The Gene Ontology \citep{Ashburner00} (GO) contains a collection
of biological concepts (GO terms) describing molecular biology
aspects of genes.  The relationship among the concepts are
represented in a directed acyclic graph (DAG).  An edge reflecting
an ``is-a'' relationship between a pair of GO terms indicates that
the concept encoded by the parent term is more general and
subsumes the concept of the child term.  The GO has been widely
used to annotate the function of genes of different model
organisms, therefore it is natural to treat a set of genes
annotated with a common GO term as a \textit{functional module}, a
widely used approach in bioinformatics analyses
\citep{SegalModuleMap2004,Subramanian2005}.

\begin{figure}[h]
\centering
 \scalebox{0.4}{\includegraphics{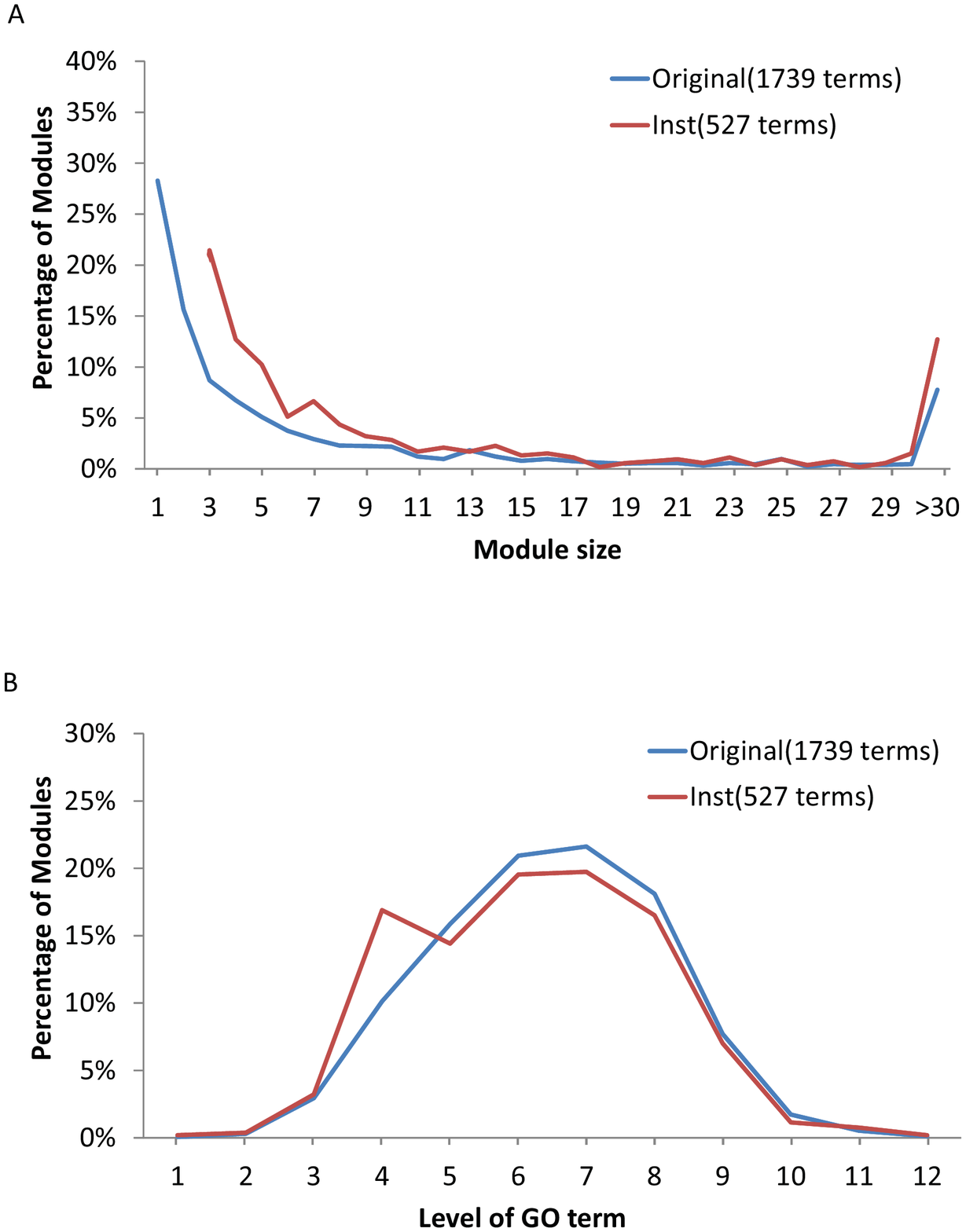}}
 \caption{{\bf  Characterization of the summary GO terms.}  {\bf
 A.}  The histograms of the number of genes associated with each
 GO term  before and after ontology-guided knowledge mining:
 1) the original GO annotations for all responding genes (blue),
 and 2) the GO terms  returned by the instance-based module search (red).  {\bf B.} The distribution of the levels  of the above GO term sets in the ontology hierarchy are shown as normalized histograms. Level $0$ represents the root of the Biological Process namespace.}
 \label{FigureHistGrams}
\end{figure}

We first investigated if original GO annotations from the GO
database are suitable to represent the major functional themes of
genes responding to perturbations in our setting.  Based on the
results of gene expression analysis performed by Hughes \textit{et
al}, $5,289$ genes  were determined to be differentially expressed
in response to one or more perturbation instance(s).  We
identified all the GO terms that have been used to annotate these
genes and retained a subset that belong to the Biological
Processes domain of the GO, which consisted of 1,739 unique GO
terms. We studied the distribution of the number of genes
annotated by each GO term, and the results are shown as a
histogram in {Figure \ref{FigureHistGrams}}. The figure shows that
a large number of original GO annotations was associated with only
a few genes; in fact almost half ($43.93\%$) of the GO terms was
associated with only 1 or 2 genes.  The results reflect the fact
that, while original  GO annotations are highly specific and
informative with regards to individual genes,  they would fail to
represent the major functional themes of a set of genes.
Therefore, there is a need to identify more general terms to
represent major functional themes.

We then formulated the task of finding functional  modules  as
follows: given a list of genes responding to a perturbation
instance and their  GO  annotations,  assign the genes into
non-disjoint functional modules, such that genes within a module
participate in  coherently related biological processes. This was
achieved by utilizing the hierarchical organization of the GO to
group a subset of genes under a suitable GO term, which retains as
much of original the semantic information as possible.    We
developed novel quantitative metrics for objectively assessing the
fitness of a summarizing GO term, which enabled us to find a term
that covered many genes and yet minimized the loss of  semantic
information the original annotations.   Our   criteria for a
summarizing GO term included: 1) requiring the summarizing term to
be statistically enriched in the input gene list, and 2) requiring
the functions of the genes in a module to be semantically coherent
when measured with a functional coherence metric previously
developed by our group (\cite{Richards2010},  see Methods
section).    This enabled us to dynamically search for suitable
terms along the GO hierarchy and to group genes under suitable
summary terms in a manner that is specific for each input gene
list, rather than using pre-fixed annotations
\citep{Subramanian2005}.  We refer to this approach as a knowledge
mining approach because it searches for a new  representation of
the function of  genes through assimilating knowledge represented
by the original annotations.

\begin{figure}[h]
\centering
 \scalebox{0.4}{\includegraphics{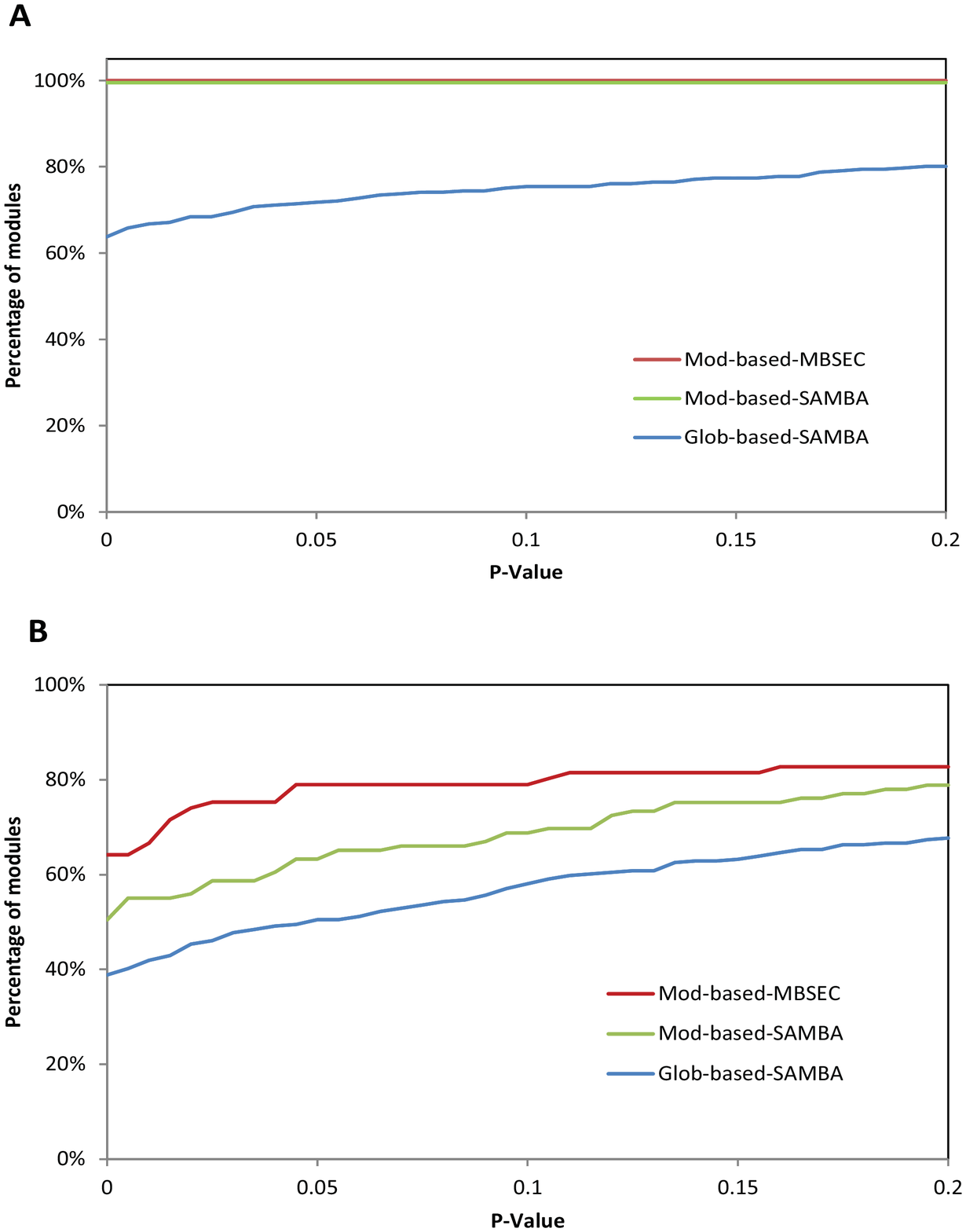}}
 \caption{
    {\bf Functional coherence of modules.}
    {\bf  A.}  The cumulative distribution of functional coherence p-values  of the
        responding modules identified by different methods:   MBSEC with module-based input graphs (red),
        SAMBA with module-based input graphs (green),
        and SAMBA with the global input graph (blue).
    {\bf B.}  The cumulative distribution of functional coherence p-values  of the
        perturbation modules identified by different methods: MBSEC with module-based input graphs (red),
        SAMBA with module-based input graphs (green),  and SAMBA with the global input graph (blue).
 }
 \label{ModuleCoherence}
\end{figure}

Applying this approach, we identified functionally coherent
modules for each perturbation experiment.  Further, we merged the
modules from different perturbation instances that shared a common
GO annotation.  The procedure led to a total of 527 distinct
functional modules, each summarized with a distinct GO term.  The
statistics of the modules, the number of genes annotated by
summarizing terms and the levels of the terms in the GO hierarchy,
are shown in Figure \ref{FigureHistGrams}. It is interesting to
note that while the summarizing GO terms tend to annotate more
genes than the original ones, the distribution of the terms along
the GO hierarchy is quite close to the original annotations,
indicating that our approach retained a level of semantic
specificity similar to the original annotations.

We further investigated the modules and found the results
biologically sensible.  For example,  we found that 38 genes were
grouped into a module annotated with the term GO:0008643 ({\it
carbohydrate transport}) (from here on, we name a functional
module using its summary GO term), including 17 genes in hexose
transport \{HXT1, HXT2, ..., HXT17\}.  The original annotations of
the genes in the module included GO:0051594 ({\it detection of
glucose}, covering 3 genes), GO:0005536 ({\it glucose binding},
covering 3 genes), GO:0005338 ({\it nucleotide-sugar transmembrane
transporter activity}, covering 4 genes), GO:0005353 ({\it
fructose transmembrane transporter activity}, covering 16 genes),
and so on.  Our algorithm summarized the function of the genes
using the term  GO:0008643 ({\it carbohydrate transport}), which
we believe does not result in a significant loss of information
regarding the individual genes, thus providing a sensible
representation of overall function of a larger group of genes.  A
list of function modules is shown in the supplementary website.

\begin{figure}[h]
\centering
 \scalebox{0.4}{\includegraphics{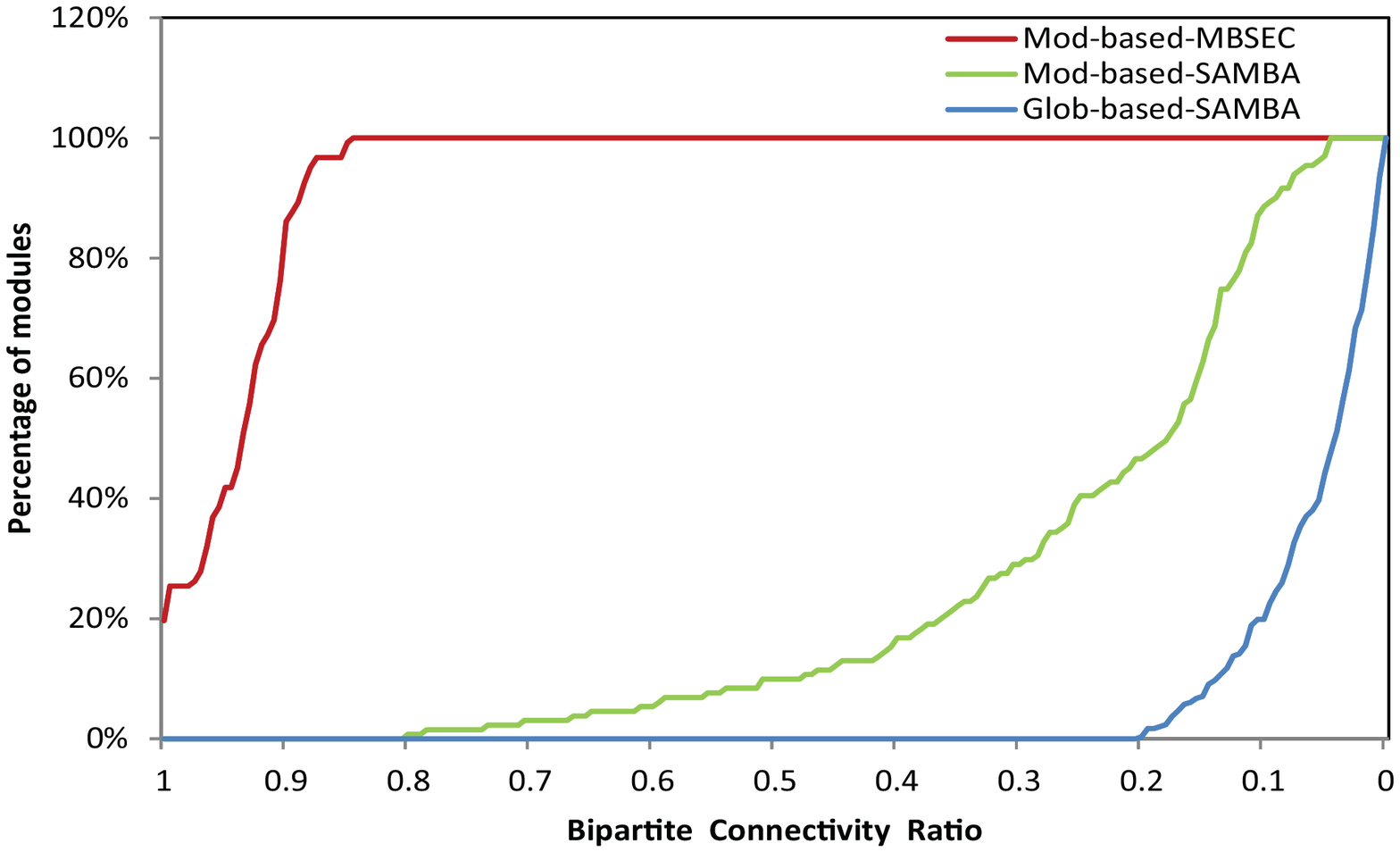}}
 \caption{
    {\bf  Subgraph connectivity.}
    Cumulative distribution of within bipartite subgraph connectivity of the modules identified in three experiments: MBSEC with module-based input graphs
     (red), SAMBA with module-based input graphs (green),  and SAMBA with global input graph (blue).
 }
  \label{WithinModuleConn}
\end{figure}

\begin{figure}[h]
\centering
 \scalebox{0.4}{\includegraphics{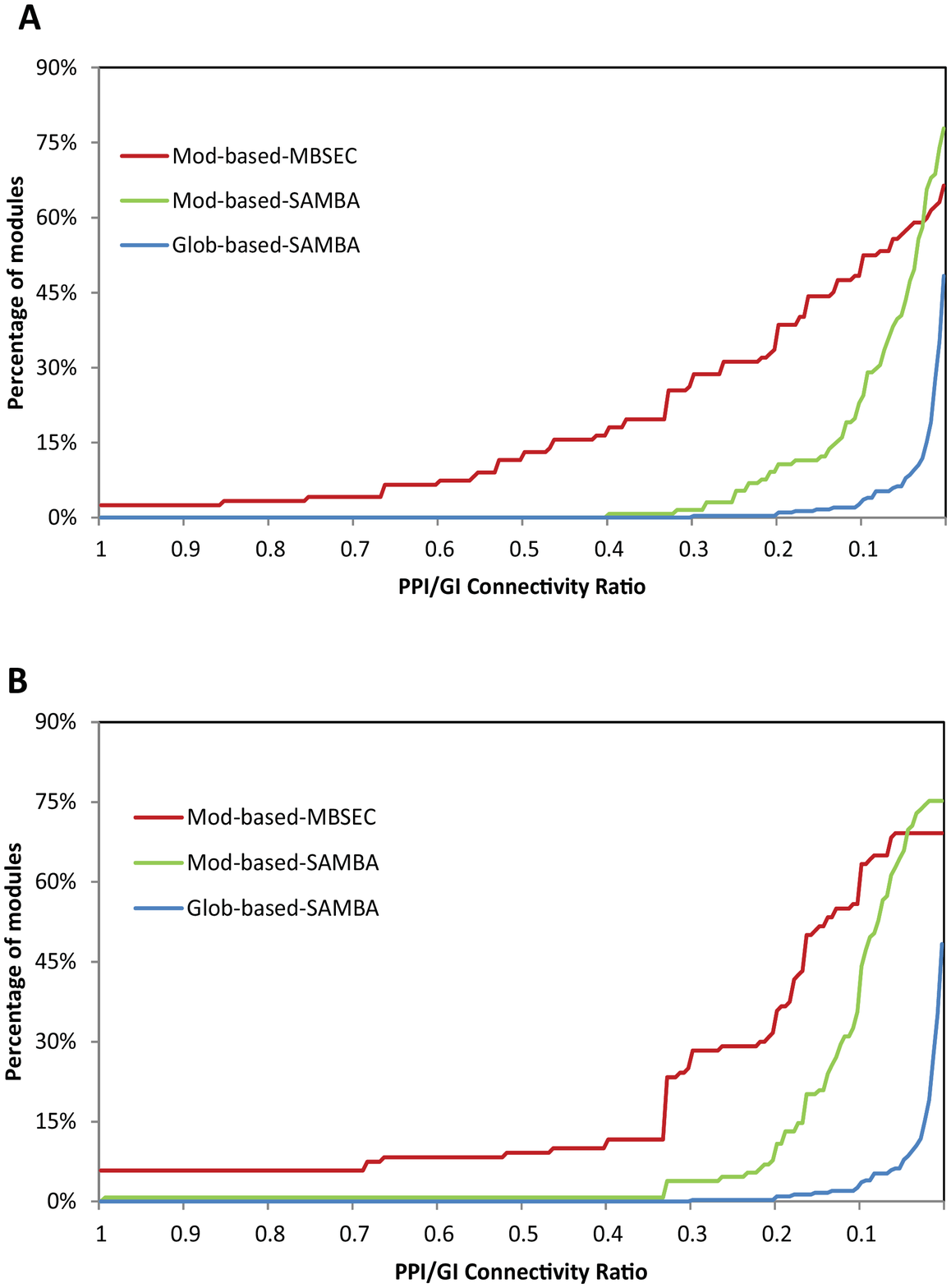}}
 \caption{
    {\bf  Protein-protein physical and genetic interactions within modules.}
    {\bf  A.}  The cumulative distribution of the within module PPI/GI connectivity ratios of  responding modules identified by different methods:
         MBSEC with module-based input graphs (red), SAMBA with module-based input graphs (green),
        and SAMBA with the global input graph (blue).
    {\bf B.}  The cumulative distribution of the connectivity ratios within
        perturbation modules identified by different methods: MBSEC with module-based input graphs (red), SAMBA with module-based
        input graphs (green),  and SAMBA with the global input graph (blue).
 }
  \label{ModuleInteraction}
\end{figure}

\subsection*{Searching for perturbation instances affecting a common signal}
Using a functional module from the previous section as a putative
unit responding to a cellular signal, we further searched for the
perturbation instances affected to the functional module.  Success
in finding a set of functionally coherent genes that repeatedly
co-responded to multiple perturbation instances  would provide a
strong indication that the responding genes are regulated as a
unit by a common signal  and that the perturbation instances may
have affected such a signal. We  addressed the searching task in
the following steps: 1) Given a functional module,  we first
created a bipartite graph using all perturbation instances on one
side and the genes in the functional module on the other side,
referred to as a functional-module-based graph.  In such a graph,
an edge between a perturbation instance and a responding gene
indicates that the gene is differentially expressed in response to
the instance. 2) We then searched for a densely connected subgraph
satisfying the following conditions: a) each vertex was on average
connected to a given percent, $r$, of the vertices on the opposite
side, and b) the size (number of vertices) of the subgraph was
maximized.  We refer to the vertices on the perturbation side of a
densely connected subgraph as a \textit{perturbation module}, and
those on the responding side as a \textit{response module}. The
problem of finding such a subgraph from a bipartite graph belongs
to a family bi-clustering problems \citep{MadeiraBiClustering2004,
Cheng2000,Erten2010}, which are NP-hard.  There are many
approximate algorithms for solving the problem (see the review by
\cite{MadeiraBiClustering2004}), but our formulation has  distinct
objectives, which allow us to specify the degree of  connectivity
between perturbation and responding modules.  We have developed
and implemented a greedy algorithm, referred to as the maximal
bipartite subgraph with expected connectivity (MBSEC) algorithm,
to solve this problem, see Methods.

We performed experiments to test the following two hypotheses: 1)
using  functional-module-based graphs as inputs for a
dense-subnetwork  searching algorithm would enhance the capability
of identifying signaling  pathways; 2) specifically pursuing high
density of a subgraphs enhances the capability of finding
signaling pathways.   To test the first hypothesis, we  applied an
algorithm referred to as the statistical-algorithmic method for
bicluster analysis (SAMBA)  by  \cite{TanaySAMBA2002} to assess
the impact of different inputs on the quality of
perturbation-response modules.  SAMBA is  a well-established
algorithm that solves the biclustering problem under a  bipartite
graph setting, which is similar to problem setting.   We first
applied the SAMBA (implemented in the Expander program, v5.2) with
default settings   to the global  bipartite graph consisting of
all 5,289 responding genes and 300 perturbations, which returned a
total of $304$ subgraphs.  We then  applied the SAMBA program to
each of the functional-module-based graphs, and a total of $131$
subgraphs were returned.   To test the second hypothesis, we
applied the MBSEC algorithms to the same functional-module-based
graphs as in the previous experiment, using the following
parameter settings: $r \geq 0.75$ and $s \geq 4$.     The
experiment identified a total of   $122$ subgraphs that  satisfied
the requirements.

We  assessed the overall quality of a perturbation (or a
responding) module by determining the functional coherence score
of the module using the method previously developed by our group
\citep{Richards2010}.  This method measures functional relatedness
of a set of genes based on the semantic similarity of their
functional annotations  and  provides a p-value of the coherence
score of a gene set.  The key idea of this method is as follows:
given a set of genes, map the genes to a weighted graph
representing the ontology structure of the GO, in which the weight
of an edge reflects the semantic distance between the concepts
represented by the a pair of GO terms; identify a Steiner tree
that connects the GO terms annotating these genes and measure how
closely the genes are located within the graph using the total
length of the tree; apply a statistical model to assess if the
genes in the set are more functionally related than those from a
random gene set.  A
 gene set with a small p-value would indicate that the functions of
the genes are coherently related to each other.

Figure \ref{ModuleCoherence} shows the results of functional
coherence analysis of responding modules (Panel A) and
perturbation modules (Panel B) by plotting the cumulative
distribution of the modules based on their p-values.   Panel A
shows that all responding modules returned by our MBSEC algorithm
and those returned by  SAMBA with functional-module-based graphs
as input were assessed as functionally coherent.  This is not
surprising, as all the input modules were functionally coherent
(p-value $\le 0.05$), and therefore the returned responding
modules, which were sets of the input modules, were likely to be
coherent.     In comparison,  when using the global
perturbation-response bipartite graph as input, about {70\%} of
the responding modules identified by SAMBA were assessed to be
coherent.  The results indicate that, while the SAMBA algorithm is
capable of identifying biclusters with coherent responding
modules, a high percentage of returned responding modules contains
a mixture of genes involved in diverse biological processes.

Since the goal is to find perturbation instances that likely
constitute a signaling pathway, it is more interesting to inspect
if the genes in a perturbation module are  coherently related. We
assessed the functional coherence of the perturbation modules
returned from the three experiments for the impact of  different
inputs and  algorithms on the results (see Panel B of Figure
\ref{ModuleCoherence}).  A higher percentage of perturbation
modules was functionally coherent when functional-module-based
graphs were used as inputs for  SAMBA when compared with those
from the SAMBA with a global graph, indicating that indeed
perturbation instances densely connected to a functionally
coherent responding module were more coherent themselves, i.e.,
they were more likely to function together.  When comparing the
results from MBSEC algorithm with those from the SAMBA, our
algorithm returned the highest percentage of functionally coherent
perturbation modules.  The results indicate that, when inputs are
the same, specifically pursuing high density subgraphs enhances
the quality of identified perturbation modules.

We further inspected the within subgraph connectivity, determined
as  the number of edges within a subgraph over the number of
maximal possible edges ($n \times m$, with $n$ and $m$
representing the number of vertices on each side respectively), to
investigate if the differences in functional coherence  of the
modules were related to the capabilities of the algorithms to find
densely connected graphs. Figure \ref{WithinModuleConn} shows that
there were striking differences in the connectivity of the
subgraphs returned from three experiments.   The results also
support the notion that enhanced capability of finding densely
connected perturbation-response bipartite graph underlies the
capability of identifying coherent modules.

In addition to assessing the functional relationship of the genes,
we further quantified and compared  within module physical and
genetic interactions, which provided another line of evidence for
assessing if genes in the modules were functionally related. Using
protein-protein physical interaction and genetic interaction data
from the BioGrid \citep{StarkBioGrid2010}, we  calculated the
ratio of the number of known interactions within a module
containing  $N$ genes over the maximum  number of possible
interactions for the module  ($1/2 * N(N-1)$).  We plot  the
cumulative distributions of modules based on their interaction
ratios in Figure \ref{ModuleInteraction}. The figure shows that
there are more physical and/or genetic interactions within both
perturbation and responding modules identified by our methods,
indicating that indeed the genes in these modules are more likely
to function together.

Taken together, these results  indicate that, by constraining the
search space to \textit{functionally coherent genes} and
explicitly requiring a degree of connectivity of subgraphs,  our
approach enhances the capability of identifying perturbation
modules in which the genes are more likely to physically interact
with each other to participate in coherently related  biological
processes.  Thus they likely participate in a common signaling
pathway and carry a common signal.

\subsection*{Discovering  signaling pathways based on perturbation-responding subgraph}
A subgraph consisting of a perturbation and a responding module
reflects the fact that the perturbation instances affected the
signal controlling the expression state of the genes in the
responding module.  It is interesting to see if a perturbation
module contains the members and/or modulators of a signaling
pathway.   Indeed, we found many of the identified perturbation
modules corresponded to well-known signaling pathways.  For
example, our analysis  identified  a subgraph consisting of a
responding module of $8$ genes  annotated by the GO term
GO:0019236 ({\it response to pheromone}) and a perturbation module
consisting of $16$  perturbation instances:
\{{\color{blue}$STE11$}, {\color{blue}$STE4$},
{\color{blue}$DIG1$}, {\color{blue}$DIG2$}, $HMG2$,
{\color{blue}$FUS3$}, $KSS1$, $RAD6$, {\color{blue}$STE7$},
{\color{blue}$STE18$}, {\color{blue}$STE5$},
{\color{blue}$CDC42$}, {\color{blue}$STE12$}, $STE24$, $SOD1$,
$ERG28$ \}.  In the list of the perturbation instances, we
highlighted (with blue font) the genes that are known to be
members of  the well-studied yeast pheromone response pathway
reviewed by \cite{Gustin}, which listed  20 gene products as the
members of the pathway. In the study by \cite{Hughes}, 12 out of
those 20 genes were deleted. We found that 10 out of these 12
perturbation instances were included in the perturbation module of
this subgraph. This result  indicates that our approach is capable
of re-constituting the majority of the genes involved in the
pheromone signaling pathway.  Inclusion of ergosterol metabolism
enzymes, ERG28 and HMG2, in the perturbation module indicates that
our approach can also identify the modulators of a signaling
pathway.

\begin{figure*}[ht]
\centering
 \scalebox{.5}{\includegraphics{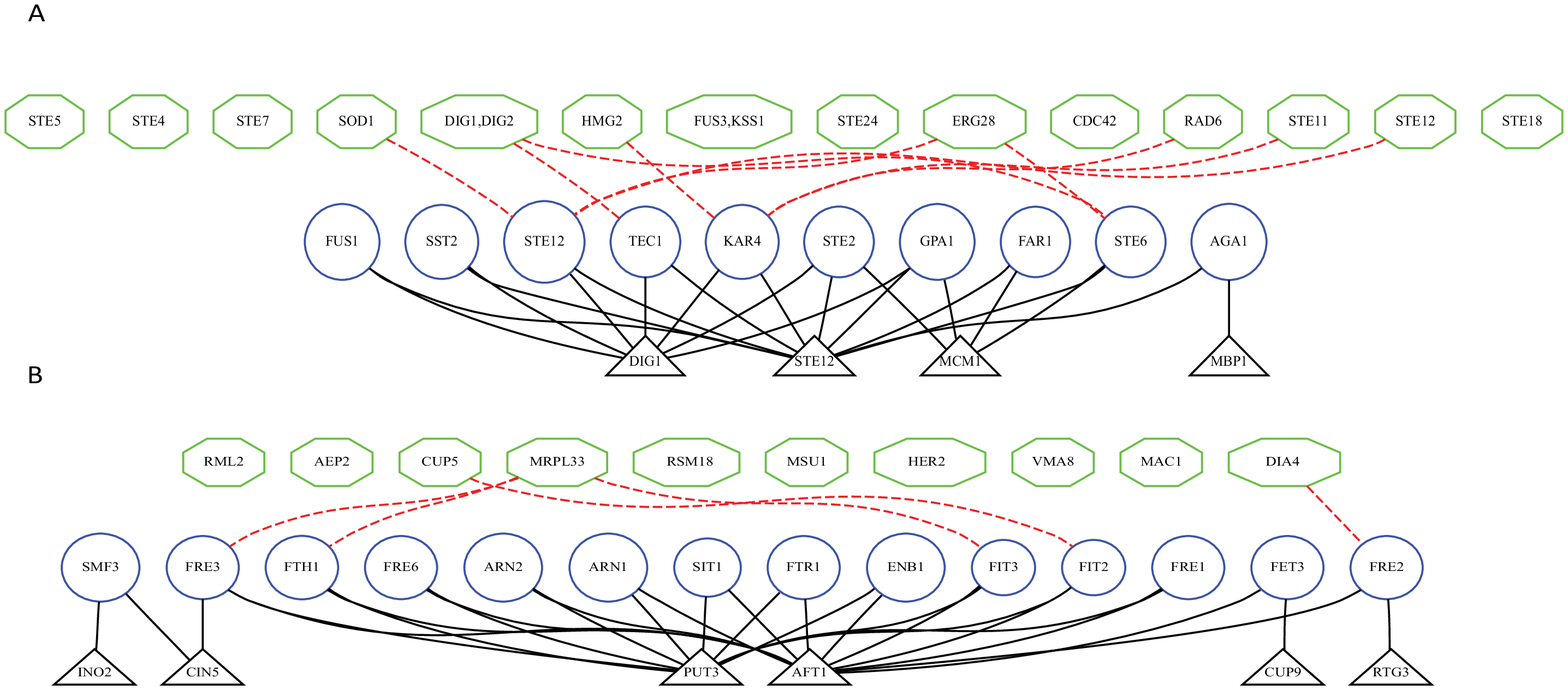}}
\caption{  {\bf  Example perturbation-responding subgraphs.}  Two
example subgraphs are shown: {\bf Panel A} GO:0019236 ({\it
response to pheromone}) and {\bf Panel B}: GO:0006826 ({\it iron
ion transport}).   For each subgraph, the  perturbation instances
(green hexagons) are shown in the top tier; responding genes (blue
circles) are shown in the middle  tiers;  and the transcription
factor modules (grey triangles)   are shown in the bottom tier. To
avoid an overly crowded figure,  a red dash line  indicates that a
perturbation instance and a responding  gene is NOT connected. }
\label{fig:PertbNetwork}
\end{figure*}

In addition to ``re-discovering'' the known signaling pathways,
analysis of subgraphs obtained in this study led novel hypotheses.
For example, in one subgraph, the responding module was annotated
with GO:0006826 ({\it iron ion transport}) and consisted entirely
of genes involved in cellular iron homeostasis, including iron
transporters and ferric reductases,  shown in  Panel B of Figure
\ref{fig:PertbNetwork}.  These genes are known to be primarily
regulated by the iron-responsive transcription factor Aft1p and
partially comprise the iron regulon in yeast
\citep{YamaguchiIwai1995}.  Intriguingly, the perturbed gene set
consisted largely of proteins involved in mitochondrial
translation, including gene products involved in mitochondrial
ribosomal subunits ($RML2$, $RSM18$, $MRPL33$), translation
($HER2$, $DIA4$, $AEP2$), and RNA processing ($MSU1$). These data
lead to a novel hypothesis that perturbation of mitochondrial
protein synthesis will lead to changes in the iron sensing
process.  In fact, such a link has only recently been suggested,
in that iron-sulfur complex synthesis in mitochondria, which
requires a set of 10 distinct protein components \citep{Lill2000},
directly impacts cellular iron uptake and utilization
\citep{Hausmann2008,Rutherford2005}. Indeed, these data provide a
rationale for the new hypothesis that mitochondria translation
plays an essential role in cell iron homeostasis through
iron-sulfur complex synthesis.

We have visualized all the perturbation-responding module pairs
identified in our experiments and show the results  on the
supplementary website.  The data allow readers, particularly yeast
biologists, to inspect the results and assess the quality of the
modules,  and more importantly, to explore new hypotheses
regarding yeast signaling systems.  In Figure
\ref{fig:PertbNetwork}, we show the  subgraphs related to
GO:0019236 ({\it response to pheromone})  and GO:0006826 ({\it
iron ion transport}).  In this figure, we show the perturbation
instances (green hexagons) and responding  modules (blue circles)
in two tiers.  Due to the fact that the connections between the
perturbation and the responding module are very dense, which would
interfere with visualization, we reversely indicate perturbation
instances and responding genes that were NOT connected,  shown as
the red dash-lines in the figure. Using a graph-based algorithm
\citep{LuS2011}, we further identified transcription factor (red
triangles) modules that are likely responsible for the
co-expression of the genes in the responding modules.  Including
TF information in data visualization further  enhances
interpretation of the subgraphs. For example, the fact that each
responding module in this figure are connected (thus potentially
regulated) by a TF module further strengthens the hypothesis that
the genes are co-regulated together as a unit responding to a
common signal.

\subsection*{Revealing organization of cellular signals}
\begin{figure*}[ht]
\centering
 \scalebox{0.9}{\includegraphics{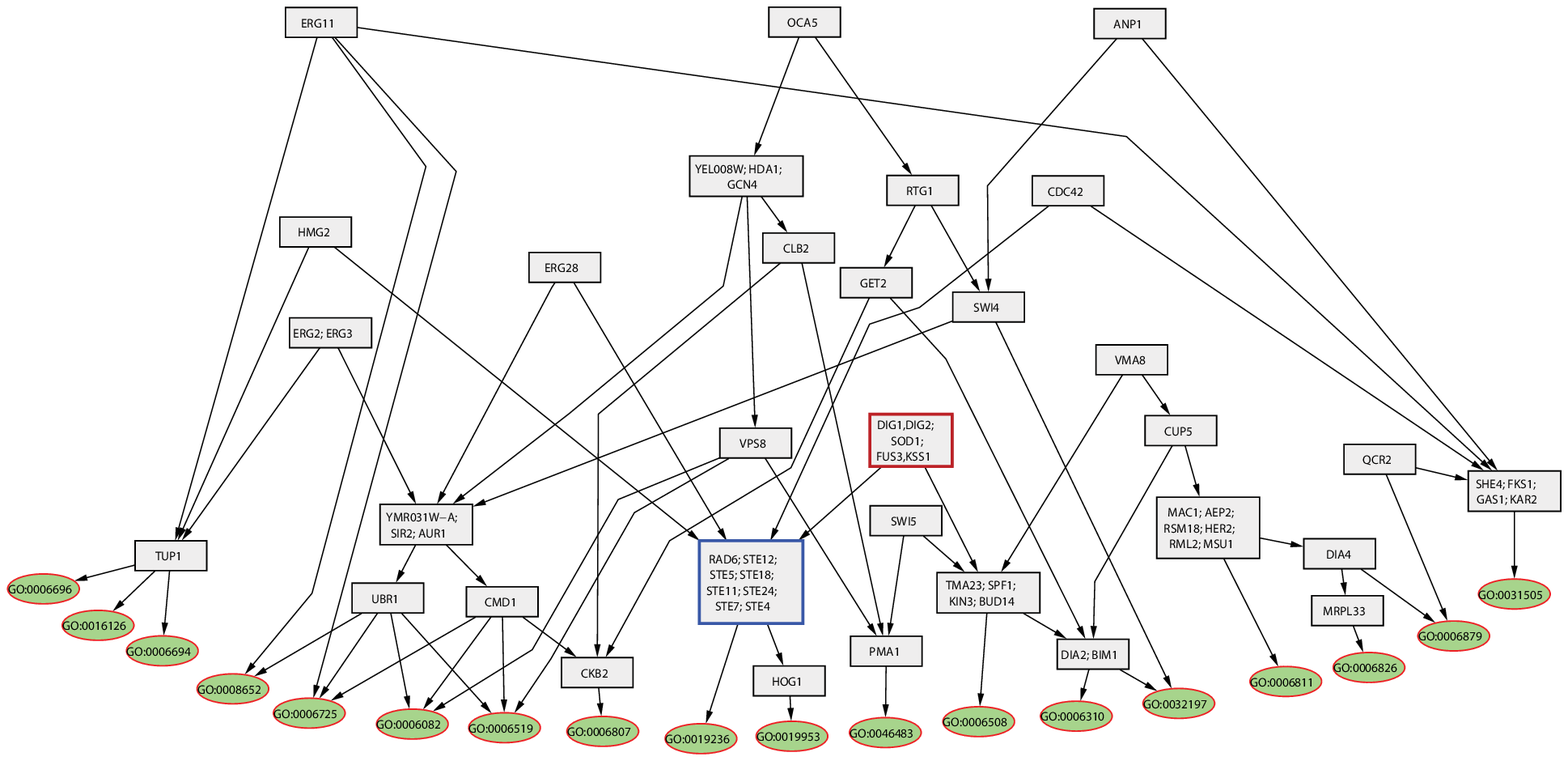}}
\caption{{\bf  Organizing perturbation instances and responding
modules}  In this graph, responding modules are represented as
green oval nodes, with each being annotated by a GO term. The
rectangle nodes are perturbation nodes, which may contain one or
more genes that share a common set of responding modules. }
\label{fig:hierachyGraph}
\end{figure*}

Our approach enabled us to use responding modules to reflect major
signals in a cellular system and  perturbation instances that
affect these signals.  We have found that many perturbation
instances were involved in multiple perturbation-response
subgraphs, indicating that the signal affected by such a perturbed
instances was connected to multiple signals through cross-talks.
This observation offered us an opportunity  to further investigate
the organization of cellular signals by studying what signals each
perturbation instance affects, and how the signals  are related to
each other.  For example, it is interesting to investigate whether
a set of perturbation instances affects a common set of responding
modules--- that is,  the information encoded by these genes is
identical---so that we can group them as a signaling unit.
Similarly, it is of interest to investigate whether the responding
modules (signals) affected by one perturbed gene are a subset of
those affected by another perturbed gene, and to utilize such a
relationship to organize the signals.  The latter task is closely
related to that addressed by the nested effect model
\citep{MarkowetzNEM2007}, which aims to capture the hierarchical
relationship among perturbation instances based on the genes they
affect.  Since the nested effect model used an individual gene as
a responding unit, the scale of the problem became intractable
(exponential) and a Markov chain Monte Carlo algorithm was
employed.  In contrast, our approach used conceptualized
responding modules, which provided two advantages: 1) the
projection of high-dimensional data at the gene level to a
low-dimensional and semantic-rich concept level reduces complexity
of the task; 2) the unique annotation associated with each module
renders the task of  determining subset relationship among
perturbation instances a trivial task.  These characteristics
enabled us to develop a polynomial algorithm (see Methods), to
organize the perturbation instances into a (DAG).  In such a
graph, each node is comprised of a  set of perturbation instances
that share common responding modules, i.e., a signaling unit; an
edge between a pair of nodes indicates that the signals affected
by the parent node subsume those carried by the child node.  We
collected all perturbation modules that contained at least $8$
perturbation instances and organized perturbation instances into a
DAG as shown in Figure \ref{fig:hierachyGraph}.

Inspecting the perturbation nodes including multiple genes, we
found that  the genes in these nodes tend to participate in
coherently related biological processes, and they often physically
interact with each other at high frequencies (data not shown). For
example, one  perturbation node  (highlighted with a blue border)
in Figure~\ref{fig:hierachyGraph} contains multiple STE
(sterility) genes,  a set of well-studied genes that mediates
pheromone signaling in yeast, and they share common responding
modules annotated with the functions ``response to pheromone''
(GO:0019236) and ``sexual reproduction'' (GO:0019953). Thus our
method is capable of identifying perturbed instances whose
information can be encoded using a one-bit signal---a switch
affecting expression of the genes responding to pheromone
signaling.

Visualization of the relationship of perturbation instances in a
DAG  enables a biologist to investigate how signals are combined
to generate a cellular response. For example, there is a
perturbation node (highlighted with a red border) in
Figure~\ref{fig:hierachyGraph} containing $DIG1$, $DIG2$, $SOD1$,
$FUS3$ and $KSS1$, all of which, except $SOD1$, are involved in
MAPK activities. Our results show that there is a  path connecting
this node to the  aforementioned STE node, and then further to the
``respond to pheromone'' responding module, indicating that the
gene products  of the two nodes  work together to transmit signals
in response to  pheromone.  Indeed, it is well known that MAPK
activities are required in the pheromone signaling pathway
\citep{Gustin,Herskowitz1995}.  Yet, our results clearly presented
the fact that the MAPK node also carries information besides
pheromone response, it also affects the biological processes of
``proteolysis'' (GO:0006508) process, for example.

Another interesting observation is that the hierarchical
organization of the perturbation instances  reflects their
relative position in a signaling cascade.  For example,
perturbation of ergosterol metabolism genes, ERG2, ERG3, HMG2,
ERG11, and ERG28, tend to have a broad impact on different
signals,  including the pheromone response pathway.   This is
understandable: as a critical component of the plasma membrane,
ergosterol influences the organizational compartments of the
plasma membrane such as lipid rafts \citep{Simon2011}, which in
turn affect the organization of signaling molecules in the
membrane. As such, perturbation of these genes has a broad impact
on diverse cellular signals.  Our results indicate that $HMG2$ and
$ERG28$ are connected to the STE node to influence the expression
of the pheromone responding module.  The role of ergosterol
metabolism in modulating pheromone response signaling has only
recently been studied by \cite{JinH2008}.   More interestingly,
our results indicate that perturbation of distinct enzymes of
ergosterol metabolism leads to distinct cellular signals,
presumably by perturbing the production of distinct species of
ergosterols.  The view that distinct lipid species encode/regulate
disparate signals is widely accepted in the lipidomics research
domain \citep{Parks1995}.

\section*{Summary}
In this study, we developed a  proof of concept framework for
unifying knowledge mining and data mining to conceptualize the
findings from systematic perturbation experiments in order  to
enhance the capability of identifying signal transduction
pathways.  The innovations of our approach are reflected in the
following aspects: 1)  an ontology-driven approach for identifying
functional modules from a genes list  in a dynamic and data-driven
(instance-based) manner and  projecting molecular finding to a
conceptual level,  2) innovative formulation of the biclustering
problem in terms of a constrained search space and new objective
functions, and 3) a novel graph algorithm that enables organizing
signaling molecules at a system level in a tractable manner for
the first time.   We have demonstrated that conceptualization of
cellular responses to systematic perturbations enhances the
capability of identifying perturbation instances that participate
in specific signal transduction pathways.   To the best of our
knowledge, this is the first report of a computational framework
capable of automatically assimilating the information from
systematic perturbation data to reveal the architecture  of a
cellular signaling system at a \textit{conceptual} level that can
be readily interpreted by biologists to gain insights of a system.

More importantly, conceptualization of experimental results is a
critical step towards the ultimate goal of systems
biology---acquiring computable knowledge from experimental data
for reasoning and hypothesis generation.  Our results already laid
the foundation to derive  abstract knowledge.  For example,   one
can translate a path from a perturbation node to a responding
module in Figure \ref{fig:hierachyGraph} into a rule as follows:
``if genes involved in MAPK signaling are perturbed, genes
involved in pheromone responses will be differentially
expressed''.   A rule like this represents the relationships
between perturbed genes and responding genes at a conceptual
level.  Equipped with rules and facts,  a computing agent can then
make a prediction that perturbation of a newly discovered gene may
lead to the differential expression of genes involved
\textit{pheromone responses}, if the gene is found to be involved
in \textit{MAPK signaling}.   Ongoing research is devoted to
acquiring and representing facts, assertions and rules from
systems biology data in an accurate and generalizable manner.

\begin{figure}[htb]
\vspace{-3mm}
\footnotesize
\begin{tabbing}
xxxx\=xx\=xx\=xx\=xx\=xx\=xx\=xx\=xx\=xx\=xx\=xx\=xx\=\kill
{\bf Algorithm-1 HDSubgraph$(G, r,s)$}\\
\textbf{Input:} $G=(V_1,V_2,E)$ -- a bipartite graph, $r$ --  the connectivity ratio of the subgraph,\\
\>\>~ and $s$ -- the minimum number of perturbations in the solution.\\

\textbf{Output:} A highly dense subgraph\\
\\

1. $G_{sub}=\emptyset$; $Score_{best}=-1$;\\
2. {\bf for each} subset $S_1$ of size $s-1$ in $V_1$ {\bf do}\\
3. \> $V_{remain} = V_1 - S_1$; $V''_1=S_1$; $Status = 1$; \\
4. \> {\bf while} $Status=1$ {\bf do}\\
5. \>\> $Score_{temp}=-1$; $G'_{sub}=\emptyset$; $Status = 0$;\\
6. \>\> {\bf for each} $u \in V_{remain}$ {\bf do}\\
7. \>\>\> $V'_1=V''_1 \cup \{u\}$; $V'_2=\{v| v \in V_2$ and $v$
connects to at least\\
\>\>\>\> $r|V'_1|$ vertices in $V'_1$\};\\
8. \>\>\> Calculate the score of induced subgraph
$G'=(V'_1,V'_2,E')$\\
\>\>\>\> and save the score to $SC$;\\
9. \>\>\> {\bf if} $SC>Score_{temp}$ {\bf then}\\
10. \>\>\>\> $Score_{temp}=SC$; $G'_{sub}=G'$;\\
11. \>\> {\bf if} $Score_{best}<Score_{temp}$ {\bf then}\\
12. \>\>\> $G_{sub}=G'_{sub}$; $Score_{best}=Score_{temp}$; $Status = 1$;\\
13. \>\>\> Assign $V'_1$ of $G_{sub}$ to $V''_1$; $V_{remain}=V_1-V''_1$;\\
14. {\bf return} $G_{sub}$;\\
{\bf Note:} 1. $score(G') = \sum_{x \in V'_1}((1+0.001)|V'_2| -
1/(1-r)(|V'_2| - degree_{G'}(x)))$. When a new  node $x$ is added, \\
\>\>\>~ there is a score gain if the degree of $x$ in $G'$ is at least $r|V'_2|$; else a penalty will be applied.\\
\>\> 2. The $s$ is smaller than or equal to the minimum number of
perturbations in the solution. The growth \\
\>\>\>~ of $s$ will greatly increase the running time of the
algorithm.

\end{tabbing}
\vspace*{-3mm}

\caption{Greedy algorithm to find the highly dense bipartite
subgraph} \label{GreedyAlg_2}
\end{figure}

\section*{Materials and Methods}
The microarray data from the systematic perturbation experiments
by \cite{Hughes} were  collected, and differentially expressed
genes responding to each perturbation were identified based the
analysis of the original paper.     Given a list of differentially
expressed genes responding to a perturbation instance, we
represent the genes and their annotations using a data structure
referred to as GOGene graph \citep{Muller09}.   In such a graph, a
node represents a GO term and a directed edge between a pair of
nodes reflects an "is-a" relationship between the GO terms; in
addition, each node keeps track of the genes it annotates,
therefore the graph contains information on both GO terms and
genes.  The procedure for searching for summarizing GO terms
iterates through the following steps:
 1) perform an enrichment analysis \citep{Khatri05} for each leaf GO term among
the instance-specific responding genes; 2) select the GO term with
the biggest p-value (least enriched) and merge its genes to the
parent node with the shortest semantic distance as defined by Jin
\textit{et al} \citep{Jin2010}; 3) trim the term off the graph; 4)
repeat the above procedures.  We stop trimming a GO term  once it
is significantly enriched (p-value $\leq$ 0.05) and the genes
summarized by the term remained functionally coherent
\citep{Richards2010}, and its associated genes are treated as a
functionally coherent module; otherwise all non-significant terms
would eventually be merged to the root node  of the GO hierarchy
and their associated genes are deemed as not coherently related.

To assess the functional coherence, we applied the method
developed by \cite{Richards2010}.  In this approach,  the ontology
structure of the GO is represented as a weighted graph, in which
an edge weight represents the semantic distances between a pair of
GO terms.  When given a list of genes,  the genes are associated
to their annotation GO terms and  a Steiner tree connecting all
genes is identified.  Using the total length of the Steiner tree
as a score reflecting the functional relatedness of the genes, a
statistical model is applied to assess the probability of
observing such a score if sets with the same size are randomly
drawn from yeast genome.   See \cite{Richards2010} for details.

To search for a densely connected perturbation-responding subgraph
in a bipartite graph, we  formulated the task as follows:  given a
bipartite graph $G$, find a subgraph $G'=(V'_1,V'_2,E')$ of $G$
that  satisfies the following conditions: 1)  $(|V'_1| \geq s )
\bigcap  (|V'_2| \geq s)$, where $s$ is a user defined threshold
for cluster size; 2)   each vertex in $V'_1$ connects to at least
$|V'_2| \times r$ vertices in $V'_2$, and each vertex in $V'_2$
connects to at least $|V'_1|\times r$ vertices in $V'_1$, where
the parameter $r\in [0, 1]$ is a connectivity ratio defined by
users; and 3) the size of the subgraph {($|V'_1| + |V'_2|$ )} is
maximized. We set the parameters as follows:  $s=4$ and $r =
0.75$.  The algorithms for searching for the subgraph is shown in
Figure \ref{GreedyAlg_2}.

\begin{figure}[htb]
\footnotesize
\begin{tabbing}
xxxx\=xx\=xx\=xx\=xx\=xx\=xx\=xx\=xx\=xx\=xx\=xx\=xx\=\kill
{\bf Algorithm organizing signaling components}\\
\textbf{Input:} A set of perturbation-responding subgraphs represented as a dictionary $D$, in which a key is \\
\>\>a perturbation instance and its value is a list of the responding modules (RMs) it connects\\

\textbf{Output:} A DAG organization of  perturbation instances and RMs\\
\\
{\it \# Create a DAG consisting of perturbation instances and RMs}\\
1.  Create an empty graph $G$;\\
2.  Add all RMs  to $G$ as RM nodes; \\
3.  Combine perturbation instances that connect  to an identical set of RMs \\
\>into a joint perturbation node;  add all resulting perturbation nodes into $G$;\\
4.  Add directed edges between perturbation nodes and RM nodes as specified in $D$\\
5.  Add directed edges between a pair of perturbation nodes,  $n_1$ and $n_2$,\\
\> if the set of RMs associated with $n_2$ is a subset of that associated with $n_1$; \\
\\

{\it \#Simplify the DAG}\\
6.  {\bf for} each node $n_1$ {\bf do}\\
6.1.\> {\bf for} each node $n_2$ that is a descendent of node $n_1$ {\bf do}\\
6.2.\>\> {\bf if} $n_2$ has a parent node that is a descendant of $n_1$ {\bf then}\\
6.3.\>\>\> Remove edge $(n_1,n_2)$;\\
7. {\bf return} $G$;
\end{tabbing}
\vspace*{-3mm}
\caption{Algorithm for organizing perturbation instances and RMs}
\label{GreedyAlg_3}
\end{figure}


To organize perturbation instances based on their signals, we
developed an algorithm to organize the perturbed instances into a
DAG. In such a graph, there are two types of nodes: responding
module nodes and perturbation nodes.  Our algorithm groups
perturbation instances that share identical responding modules
into a common perturbation node, a signaling unit, and connect the
perturbation node to its corresponding responding modules.  The
algorithm further  organizes perturbation nodes such that, if
signals by a perturbation node subsume those of another, a
directed edge pointing to the subsumed node is added between them.
The algorithm is shown in Figure \ref{GreedyAlg_3}.

\section*{Acknowledgement}
The authors would like to thank Ms Vicky Chen and Joyeeta
Dutta-Moscato for reading and editing manuscript, and Drs. Nabil
Matmati  and David Montefusco for discussions.

\paragraph{\it Funding:} This research was partially supported by the
following NIH grants:  R01LM011155 and R01LM010144.

\section*{Author Contribution}
XL conceived the project; SL performed the majority of the
analyses; BJ contributed to the methods of knowledge mining; LAC
helped with biological interpretation of results; XL, SL and LAC
drafted the manuscript.


\end{document}